\documentclass[
  ,draft            
  ,numberedheadings 
  ]
  {aipproc}

\layoutstyle{6x9}

\def\beq{\begin{equation}}
\def\bea{\begin{eqnarray}}
\def\eeq{\end{equation}}
\def\eea{\end{eqnarray}}

\begin{document}

\title{The supersymmetric standard model from the $\mathbf{Z}_6'$ orientifold?}


\author{David Bailin}{
  address={Department of Physics \& Astronomy, University of Sussex, Brighton, BN1 9QH, UK},
  ,email={d.bailin@sussex.ac.uk}
}

\author{Alex Love}{
  address={Department of Physics \& Astronomy, University of Sussex, Brighton, BN1 9QH, UK}
}


\begin{abstract}
 We construct  ${\mathcal N}=1$ supersymmetric fractional branes on the $\mathbf{Z}_6'$ orientifold. 
Intersecting stacks of such  branes are needed to build a supersymmetric standard model.
   If $a,b$  are the   stacks 
  that generate the  $SU(3)_c$ and $SU(2)_L$ gauge particles,  
 then,  in order to obtain {\em just} the chiral spectrum of the (supersymmetric)
  standard model (with non-zero Yukawa couplings to the Higgs mutiplets),
   it is necessary that  the number of intersections $a \cap b$ of the stacks $a$ and $b$, and 
  the number of intersections $a \cap b'$ of $a$ with the orientifold image $b'$ of $b$
   satisfy $(a \cap b,a \cap b')=(2,1)$ or $(1,2)$. 
It is also necessary that there is no matter in symmetric representations of the gauge group.
 We have found a number of examples having these properties. Different lattices give different solutions 
and different physics.
 
\end{abstract}

\maketitle


\section{Introduction}
Intersecting D-branes provide an attractive, bottom-up route to standard-like model building \cite{Lust:2004ks}. 
In these models one starts with 
 two stacks, $a$ and $b$ with $N_a=3$ 
and $N_b=2$, of D6-branes wrapping the three large spatial 
dimensions plus 3-cycles of the six-dimensional  internal space (typically a torus $T^6$ 
or a Calabi-Yau 3-fold) on which the theory is compactified.
 These generate  the gauge group $U(3) \times U(2) \supset SU(3) _c \times SU(2)_L$, and  the non-abelian component of the standard model gauge group
is immediately assured.
   Further, (four-dimensional) fermions in bifundamental representations 
$({\bf N} _a, \overline{\bf N}_b)= ({\bf 3}, \overline{\bf 2})$ 
of the gauge group can arise at the multiple intersections of the two stacks. 
These are precisely the representations needed for the quark doublets $Q_L$ of the Standard Model.
 In general, intersecting branes yield a non-supersymmetric spectrum, so that, to avoid the hierarchy problem, the string scale associated 
 with such models must be low, no more than a few TeV. Then, the high energy (Planck)  scale associated with gravitation 
does not emerge naturally. Nevertheless, it seems that these problems can be surmounted \cite{Blumenhagen:2002vp,Uranga:2002pg}, and indeed an 
attractive model having just the spectrum of the standard model has been constructed \cite{Ibanez:2001nd}. It uses D6-branes that wrap 3-cycles 
of an orientifold $T^6/\Omega$, where $\Omega$ is the world-sheet parity operator. The advantage and, indeed, the necessity of using 
an orientifold stems from the fact that for every stack $a,b, ...$ there is an orientifold image $a',b', ...$. 
At intersections of $a$ and $b$ there are chiral fermions 
in the $({\bf 3}, \overline{\bf 2})$ representation of $U(3) \times U(2)$, where the ${\bf 3}$ has charge $Q_a=+1$ with respect to the 
$U(1)_a$ in $U(3)=SU(3)_c \times U(1)_a$, and the $\overline{\bf 2}$ has charge $Q_b=-1$ with respect to the 
$U(1)_b$ in $U(2)=SU(2)_L \times U(1)_b$.  However, at intersections of $a$ and $b'$ there are chiral fermions 
in the $({\bf 3},{\bf 2})$ representation, where  the ${\bf 2}$ has $U(1)_b$ charge $Q_b=+1$. In general, besides gauge bosons, 
stacks of D-branes on orientifolds also have chiral matter in the 
symmetric $\mathbf{S}$ and antisymmetric $\mathbf{A}$ representations of the relevant gauge group; 
both have charge $Q=2$ with respect to the relevant $U(1)$. For the stack $a$ with $N_a=3$, 
 $\mathbf{S}_a=\mathbf{6}$ and  $\mathbf{A}_a=\overline{\mathbf{3}}$. 
The former must be excluded on phenomenological grounds, but the latter could be quark-singlet states $q^c_L$. Similarly, 
for the stack $b$ with $N_b=2$, 
 $\mathbf{S}_b=\mathbf{3}$ and  $\mathbf{A}_b=\mathbf{1}$. 
Again, the former must be excluded on phenomenological grounds, but the latter could be lepton-singlet states $\ell^c_L$. 
Suppose that the number of intersections $a \cap b$ of the stack $a$ with $b$ is $p$, 
the number of intersections $a \cap b'$ of the stack $a$ with $b'$ is $q$, and the number of copies of 
$\mathbf{A}_a=\overline{\mathbf{3}}$ is $r$. The standard model has 3 quark doublets $Q_L$, 
so that to get just the standard-model spectrum we must have $p+q=3$. 
The standard model also has  a total of 6 quark-singlet states.  
To get just the standard model spectrum we also require that $6-r$ of the quark singlets 
 arise from intersections of $a$ with other stacks 
$c,d,...$ having just a single D6-brane. These belong to the representation $({\bf 1}, \overline{\bf 3})$ of 
$U(1) \times U(3)$ and each has charge $Q_a=-1$.  Ramond-Ramond (RR) tadpole cancellation requires that overall $Q_a$ sums 
to zero.
Thus
\beq
2p+2q+2r-(6-r)=0
\eeq
Hence $r=0$ and we must also exclude the representations $\mathbf{A}_a=\overline{\mathbf{3}}$. 
Tadpole cancellation also requires that $Q_b$ sums to zero overall. To get just the standard model spectrum we require that there 
are 3 lepton doublets $L$ arising from intersections of $b$ with other stacks  having just a single D6-brane. All have 
$Q_b=+1$ or $Q_b=-1$. Suppose the number of copies of 
$\mathbf{A}_b=\mathbf{3}$ is $s$.
Then overall cancellation of $Q_b$ requires that
\beq
-3p+3q +2s \pm 3=0
\eeq
Hence $s=0 \bmod 3$. In the case that $s=0$ the solutions are $(p,q)=(1,2)$ or $(2,1)$, whereas when $s=\pm 3$ the 
solutions $(p,q)=(3,0)$ or $(0,3)$ are also allowed \cite{Blumenhagen:2001te}. (Models with $|s|>3$ will obviously have non-standard model spectra.) 
However, states arising as the antisymmetric representation of $U(2)$ do not have the standard-model Yukawa
 couplings to the Higgs multiplet. Consequently we are only interested in models such as that in  \cite{Ibanez:2001nd} with 
 $(a \cap b ,a \cap b')=(1,2)$ or $(2,1)$.

Despite the attractiveness of that model, there remain serious problems in the absence of supersymmetry.
 A generic feature  of intersecting brane models 
is that flavour changing neutral currents 
are generated by four-fermion operators induced by string instantons \cite{Abel:2003yh}. The severe experimental limits on these processes 
require that the string scale is rather high, of order $10^4$ TeV. This makes the fine tuning problem very severe, and the viability 
of such models highly questionable. Further, in  non-supersymmetric theories, such as these, the cancellation of RR tadpoles does not ensure 
Neveu Schwarz-Neveu Schwarz (NSNS) tadpole cancellation. NSNS tadpoles are simply the first
 derivative of the scalar potential with respect to the scalar fields, specifically the complex structure  and K\"ahler moduli 
 and the dilaton. A non-vanishing derivative of the scalar potential signifies that  
 such scalar fields are not even solutions of the equations of motion. 
 Thus a particular consequence of the non-cancellation is that the complex structure moduli are unstable \cite{Blumenhagen:2001mb}. 
 One way to stabilise these moduli 
  is for the 
D6-branes to wrap 3-cycles of an orbifold $T^6/P$, where $P$ is a point group, rather than a torus $T^6$. 
The FCNC problem can be solved and the complex structure moduli stabilised when the theory is supersymmetric. 
First, a supersymmetric theory is not obliged to have the low string scale that led to problematic FCNCs  induced by string instantons. 
Second, 
 in a supersymmetric theory, RR tadpole cancellation ensures cancellation 
of the NSNS tadpoles \cite{Cvetic:2001tj,Cvetic:2001nr}.
An orientifold is then constructed by quotienting the orbifold with the world-sheet parity operator $\Omega$.
(
As explained above, an orientifold is necessary to allow the possibility of obtaining just the spectrum of the supersymmetric standard model.)

Several attempts  have been made to construct the MSSM \cite{Blumenhagen:2002gw, Honecker:2003vq,
Honecker:2004np,
Honecker:2004kb} using 
 an orientifold with point group  $P=\mathbf{Z}_4$, $\mathbf{Z}_4 \times \mathbf{Z}_2$ or $\mathbf{Z}_6$. 
 The most successful attempt to date is the last of these \cite{Honecker:2004kb, Ott:2005sa},
 which uses D6-branes intersecting on  a $\mathbf{Z}_6$ orientifold
 to construct an $\mathcal{N}=1$ supersymmetric standard-like model using 5 stacks of branes. 
We shall not discuss this beautiful model in any detail except to note that the intersection numbers for the stacks $a$, 
which generates the $SU(3)_c$ group, and $b$, which generates the $SU(2)_L$, are  $(a \cap b,a \cap b')=(0,3)$.
 In this case it is impossible to obtain lepton singlet states $\ell ^c_L$ as antisymmetric representations of $U(2)$. 
Further, it was shown, quite 
generally, that it is impossible to find stacks $a$ and $b$ such that $(a \cap b,a \cap b')=(2,1)$ or $(1,2)$. 
 Thus, as 
explained above, it is impossible to obtain exactly the spectrum of the (supersymmetric) standard model. 

The question then arises as to whether the use of a different orientifold could circumvent this problem. Here we 
address this question for the $\mathbf{Z}_6'$ orientifold. We do not attempt to construct a standard(-like) MSSM. Instead, we merely see 
whether there are any stacks $a,b$ that simultaneously satisfy the supersymmetry constraints, the absence of chiral matter in symmetric 
 representations of the gauge groups (see below), 
 which have not too much chiral matter in antisymmetric representations of the gauge groups,
  and which have $(a \cap b,a \cap b')=(2,1)$ or $(1,2)$.
Further details of this work may be found in reference \cite{Bailin:2006zf}.
   
\section{$\mathbf{Z}_6'$ orientifold}
We assume that the torus $T^6$ factorises into three 2-tori $T^2_1 \times T^2_2 \times T^2_3$. 
The  2-tori $T^2_k \ (k=1,2,3)$ are parametrised by complex coordinates  $z_k$. 
 The action of the generator $\theta$ of the point group  $\mathbf{Z} ' _6$ on the   
coordinates $z_k $ is given by
\beq
\theta z_k = e^{2\pi i v_k} z_k
\eeq
where
\beq
 (v_1,v_2,v_3)
= \frac{1}{6} (1,2,-3) 
 \label{z61vk}
\eeq
The point group action must be an automorphism of the lattice, so 
in $T^2_{1,2}$ we may take an $SU(3)$ lattice. 
Specifically we define the basis 1-cycles 
 by $\pi _1$ and $\pi _2 \equiv e^{i\pi /3} \pi _1$ in $T^2_1$, and 
$\pi_3$ and 
$\pi _4 \equiv e^{i\pi /3} \pi _3$ in $T^2_2$. Thus the complex structure of these tori is given by 
$U_1=e^{i\pi /3}=U_2$. The orientation of $\pi _{1,3}$ relative to the real 
and imaginary axes of $z_{1,2}$ is arbitrary. 
 Since $\theta $ acts as a reflection in $T^2_3$, the lattice, with basis 1-cycles $\pi _5$ and $\pi _6$, 
 is arbitrary. The point group action on the basis 1-cycles is then 
\bea
\theta \pi _1 = \pi _2 \quad &{\rm and}& \quad \theta \pi _2 =\pi _2-\pi _1  \label{theta12} \\
\theta \pi _3  =\pi _4 -\pi _3 \quad &{\rm and}& \quad \theta \pi _4 =-\pi _3  \label{theta34} \\
\theta \pi _5=-\pi _5 \quad &{\rm and}& \quad \theta \pi _6 =-\pi _6 \label{thetaZ6}
\eea 
We consider  ``bulk'' 3-cycles of $T^6$  which are linear combinations of the 8 3-cycles
$\pi_{i,j,k} \equiv \pi _i \otimes \pi _j \otimes \pi _k$ where $i=1,2, \ j=3,4, \ k=5,6$. The basis of 3-cycles that are 
{\em invariant} under the action of $\theta$ contains 4 elements $\rho _{1,3,4,6}$, where 
\bea
\rho _1 &=& 2(\pi _{1,3,5} + \pi _{2,3,5} +\pi _{1,4,5} - 2 \pi_{2,4,5}) \label{rho1} \\
\rho _3 &=& 2(-2\pi _{1,3,5} + \pi _{2,3,5} +\pi _{1,4,5} +  \pi_{2,4,5}) \label{rho3} 
\eea
and similarly for $\rho _{4,6}$ replacing $\pi _5$ by $ \pi _6$ in $\rho_{1,3}$ respectively.
Then the general $\mathbf{Z}_6'$-invariant bulk 3-cycle 
with (co-prime) wrapping numbers $(n_k,m_k)$ of  the cycles $(\pi _{2k-1},\pi _{2k})$ on $T^2_k$ is 
\beq
\Pi _a=A_1\rho _1+ A_3 \rho_3 +A_4\rho _4+ A_6 \rho_6 \label{genbulk}
\eeq
where 
\bea 
A_1= (n_1n_2+n_1m_2+ m_1n_2)n_3  \label{A1}\\
A_3= (m_1m_2+n_1m_2+ m_1n_2)n_3 \\
A_4= (n_1n_2+n_1m_2+ m_1n_2)m_3 \\
A_6= (m_1m_2+n_1m_2+ m_1n_2)m_3 \label{A6}
\eea 
are the ``bulk coefficients''.  
If $\Pi _a$ has wrapping numbers $(n^a_k, m^a_k)$ $(k=1,2,3)$, and $\Pi _b$ has wrapping numbers $(n^b_k, m^b_k)$, then, 
in an obvious notation,  
the intersection number of the orbifold-invariant 3-cycles  is
\bea
\Pi _a \cap \Pi _b = -4(A^a_1A^b_4-A^a_4A^b_1)&+& 2(A^a_1A^b_6-A^a_6A^b_1) +2(A^a_3A^b_4-A^a_4A^b_3) - \nonumber \\
 &-&4(A^a_3A^b_6-A^a_6A^b_3)       \label{pia0pib}
\eea
which is always even.  

Besides these (untwisted) 3-cycles, there are also exceptional 3-cycles associated with (some of) the 
twisted sectors of the orbifold. They arise in twisted sectors in which there is a fixed torus, 
and consist of a collapsed 2-cycle at a fixed point times a 1-cycle in the invariant plane.
We shall only be concerned with those that arise in the 
$\theta ^3$ sector, which has $T^2_2$ as the invariant plane. 
There is a $\mathbf{Z}_2$ symmetry 
acting in $T^2_1$ and $T^2_3$ and this has sixteen fixed points $f_{i,j}$ where $i,j=1,4,5,6$. 
There are then 32 independent exceptional cycles given by $f_{i,j} \otimes  \pi _{3,4}$ from which 8
 independent $\mathbf{Z}_6'$-invariant 
combinations may be formed. They are
\bea
\epsilon _j\equiv (f_{6,j}-f_{4,j}) \otimes 
\pi _3+ (f_{4,j}-f_{5,j}) \otimes \pi _4    \label{epsj}\\
 \tilde{\epsilon} _j\equiv(f_{4,j}-f_{5,j}) \otimes \pi _3+ (f_{5,j}-f_{6,j}) \otimes \pi _4 
  \label{epstilj}
  \eea
The non-zero intersection numbers for the  invariant combinations  are given by
\beq
\epsilon _j \cap \tilde{\epsilon} _k=-2 \delta _{jk} \label{eps0jk}
\eeq
and again these are always even.
The relation between the fixed points $f_{i,j}$ and the 
invariant exceptional cycles is given in Table \ref{FPex}
\begin{table}
\begin{tabular}{cc} \hline 
\tablehead{1}{c}{b}{Fixed point $\otimes$ 1-cycle} & \tablehead{1}{c}{b}{Invariant exceptional 3-cycle} \\  \hline
$f_{1,j} \otimes (n_2 \pi _3 +m_2 \pi _4)$ & 0 \\ 
$f_{4,j} \otimes (n_2 \pi _3 +m_2 \pi _4)$ & $m_2 \epsilon _j + (n_2+m_2) \tilde{\epsilon _j}$ \\ 
$f_{5,j} \otimes (n_2 \pi _3 +m_2 \pi _4)$ & $-(n_2+m_2) \epsilon _j-n_2 \tilde{\epsilon _j} $ \\ 
$f_{6,j} \otimes (n_2 \pi _3 +m_2 \pi _4)$ & $n_2 \epsilon _j-m_2 \tilde{\epsilon _j} $ \\ \hline 
\end{tabular}
\caption{ \label{FPex} Relation between fixed points and exceptional 3-cycles.}
\end{table}

The embedding $\mathcal{R}$ of the world-sheet parity operator $\Omega$ 
 may be chosen to act on the three complex coordinates $z_k \ (k=1,2,3)$ as
complex conjugation 
$\mathcal{R}z_k=\overline{z} _k$,
and we require that this too is an automorphism of the lattice. 
This fixes the orientation of the basis 1-cycles in each torus relative to the 
Re $z_k$ axis. It requires them to be in one of two configurations {\bf A} or {\bf B}. 
When $T^2_1$ is in the {\bf A} configuration, $\pi _1$ is aligned along the Re $z_1$ axis, whereas in the 
{\bf B} configuration it makes an angle of $\pi /6$ below this axis. Similarly for $\pi _3$ and $T^2_2$. 
In $T^2_3$  the cycle $\pi _5$ is  is aligned along the Re $z_3$ axis
in both {\bf A} and {\bf B} configurations. 
The difference is that in {\bf A} the 1-cycle $\pi _6$  aligned along the Im $z_3$ axis, 
whereas in {\bf B} it is inclined such that its real part is one half that of $\pi _5$. In 
both cases the imaginary part is arbitrary, and so therefore  is the imaginary part of the complex structure $U_3$ of $T^2_3$.
It is then straightforward to determine the action of $\mathcal{R}$ on the bulk 3-cycles $\rho _p\ (p=1,3,4,6)$ 
and on the exceptional cycles $\epsilon _j$ and $\tilde{\epsilon} _j$. In particular, requiring that a bulk 3-cycle 
$\Pi _a = \sum _p A_p \rho _p$ 
be invariant under the action of $\mathcal{R}$ gives 2 constraints on the bulk coefficients $A_p$, so that just 2 
of the 4 independent bulk 3-cycles are $\mathcal{R}$-invariant. Which 2 depends upon the lattice.

 The twist (\ref{z61vk}) ensures that the closed-string sector is supersymmetric.
  In order to avoid supersymmetry breaking in the open-string sector, 
the D6-branes must wrap special Lagrangian cycles. Then the  stack $ \Pi _a$
 with wrapping numbers $(n^a_k,m^a_k) \ (k=1,2,3)$ is supersymmetric if 
\beq
\sum _{k=1}^3 \phi^a _k= 0 \bmod 2\pi \label{sumphik}
\eeq
where $\phi^a _k$ is the angle that the 1-cycle in $T^2_k$ makes with the Re $z_k$ axis.
Defining
\beq
Z^a \equiv \prod _{k=1}^3\pi_{2k-1}(n^a_k+m^a_kU_k) \equiv X^a+iY^a
\eeq
where $U_k$ is the complex structure on $T^2_k$, 
the condition (\ref{sumphik}) that $\Pi_a$ is  supersymmetric may be written as
\beq
X^a>0, \  Y^a=0 \label{XaYa}
\eeq
(A stack with $Y^a=0$ but $X^a<0$, so that $\sum _k \phi^a _k= \pi \bmod 2\pi$, corresponds to a (supersymmetric) stack of anti-D-branes.)   
In our case $T^2_{1,2}$ are $SU(3)$ lattices, and $U_1=e^{i\pi /3}=U_2$, as already noted . Thus
\beq
 Z^a= \pi_1\pi_3\pi_5[A^a_1-A^a_3+U_3(A^a_4-A^a_6)+e^{i\pi /3}(A^a_3+A^a_6U_3)]
 \eeq
 It is 
then straightforward 
to evaluate  $X^a$ and $Y^a$ for the different lattices.
The results  for the cases in which $T^2_3$ is of {\bf B} type are given in Table \ref{susy3cycle}.
\begin{table}
\begin{tabular}{ccc} \hline 
\tablehead{1}{c}{b}{Lattice} & \tablehead{1}{c}{b}{ $X^a$} & \tablehead{1}{c}{b}{$Y^a$} \\ \hline 
{\bf AAB}&$ 2A^a_1-A^a_3+A^a_4-\frac{1}{2}A^a_6-A^a_6\sqrt{3}{\rm Im} \ U_3 $  & $\sqrt{3}(A^a_3+\frac{1}{2}A^a_6)+(2A^a_4-A^a_6){\rm Im} \ U_3 $ \\ 
{\bf ABB} and {\bf BAB} &$\sqrt{3}(A^a_1+\frac{1}{2}A^a_4)+(A^a_4-2A^a_6){\rm Im} \ U_3 $& $2A^a_3-A^a_1+A^a_6-\frac{1}{2}A^a_4+A^a_4\sqrt{3}{\rm Im} \ U_3$ \\ 
{\bf BBB} &$(A^a_3+A^a_1+\frac{1}{2}A^a_6+\frac{1}{2}A^a_4)+$& $\sqrt{3}(A^a_3-A^a_1+\frac{1}{2}A^a_6-\frac{1}{2}A^a_4)$ \\ 
&$+(A_4-A_6)\sqrt{3}{\rm Im} \ U_3 $ & $+(A_4+A_6){\rm Im} \ U_3$ \\ \hline 
\end{tabular}
\caption{ \label{susy3cycle} The functions $X^a$ and $Y^a$. (An overall  positive factor is 
omitted.)  A  stack $a$ of 
D6-branes is supersymmetric if  $X^a>0$ and $Y^a=0$.}
 \end{table}
 The (single) requirement that $Y_a=0$ means that 3 independent combinations of the 4  invariant bulk 3-cycles may be chosen to be 
 supersymmetric. Of these, 2 are the $\mathcal{R}$-invariant combinations. However, 
 unlike in the case of the $\mathbf{Z}_6$ orientifold, in this case there is a third, independent, supersymmetric bulk
  3-cycle that is {\em not} 
 $\mathcal{R}$-invariant.  
 
We  noted earlier that the intersection numbers of both the bulk 3-cycles $\rho _p \ (p=1,3,4,6)$ and of the 
exceptional cycles $\epsilon _j, \tilde{\epsilon} _j \ (j=1,4,5,6)$ are always even. However, in order to get just the (supersymmetric)
 standard-model spectrum,  either $a \cap b$ or $a \cap b'$ must be odd. It is therefore necessary to use fractional branes 
of the form
\beq
a= \frac{1}{2} \Pi _a^{\rm bulk}+ \frac{1}{2} \Pi _a^{\rm ex} \label{pifrac}
\eeq
where $\Pi _a^{\rm bulk}=\sum _p A_p \rho _p$ is an invariant bulk 3-cycle, 
 associated with wrapping numbers $(n^a_1,m^a_1)(n^a_2,m^a_2)(n^a_3,m^a_3)$,
 as shown in (\ref{genbulk}). The 
exceptional branes (in the $\theta ^3$ sector) are associated with the fixed points $f_{i,j}, \ (i,j=1,4,5,6)$ in $T^2_1 \otimes T^2_3$,
 as shown in 
(\ref{epsj}) and (\ref{epstilj}). If $\Pi _a^{\rm bulk}$ is a supersymmetric bulk 3-cycle, then the fractional brane $a$,
 defined in (\ref{pifrac}), preserves supersymmetry 
provided that the exceptional part  $\Pi _a ^{\rm ex}$ arises only from fixed points traversed by the bulk 3-cycle. 
Since  the wrapping numbers $(n^a_1,m^a_1)$ on $T^2_1$ are integers, the 1-cycle on $T^2_1$ either traverses zero fixed points or 
two. In the latter case we denote the fixed points by $(i^a_1,i^a_2)$.
  Similarly for the 1-cycle on $T^2_3$, where the 
two fixed points are denoted by $(j^a_1,j^a_2)$. 
Thus, supersymmetry requires that  the exceptional part $\Pi _a^{\rm ex}$ of $a$ derives from  four fixed points,  
$f_{i^a_1j^a_1},f_{i^a_1j^a_2},f_{i^a_2j^a_1},f_{i^a_2j^a_2}$. 
 The choice of Wilson lines affects the relative signs with which the contributions from the four fixed points are combined 
 to determine $\Pi _a ^{\rm ex}$. The rule is that
 \beq
 (i^a_1,i^a_2)(j^a_1,j^a_2)
 \rightarrow (-1)^{\tau^a _0} \left[ f_{i^a_1j^a_1}+(-1)^{\tau^a _2}f_{i^a_1j^a_2}+(-1)^{\tau^a _1}f_{i^a_2j^a_1}
 +(-1)^{\tau^a _1+\tau^a _2}f_{i^a_2j^a_2} \right]
 \eeq
where $\tau^a _{0,1,2}=0,1$ with $\tau^a _1 =1$ corresponding to a Wilson line in $T^2_1$ and likewise for $\tau^a _2$ in $T^2_3$. 
The fixed point $f_{i^a,j^a}$   with 1-cycle $n^a_2\pi _3+m^a_2\pi _3$ is then associated with the orbifold invariant 
exceptional cycle as shown in Table \ref{FPex}.

 In general, besides the chiral matter  in bifundamental representations that occurs at the intersections of brane stacks $a,b,...$,
    with each other or with their orientifold images $a', b',...$, there is also  
 chiral matter in the symmetric ${\bf S}_a$ and antisymmetric representations ${\bf A}_a$ of the gauge group $U(N_a)$, and 
 likewise for $U(N_b)$. Orientifolding induces topological defects, O6-planes, which are sources of RR charge. 
The number of multiplets in the ${\bf S}_a$ and ${\bf A}_a$ representations is 
\bea
\#({\bf S}_a)=\frac{1}{2}(a \cap a' -a \cap \Pi _{\rm O6}) \\
\#({\bf A}_a)=\frac{1}{2}(a \cap a' +a \cap \Pi _{\rm O6})
\eea
 where 
$\Pi _{\rm O6}$ is the total O6-brane homology class; it is $\mathcal{R}$-invariant. 
If $a \cap \Pi _{\rm O6}=\frac{1}{2} \Pi _a ^{\rm bulk} \cap \Pi _{\rm O6}\neq 0$, 
then copies of one or both representations are inevitably present. Since we require supersymmetry, $\Pi _a ^{\rm bulk}$ is necessarily 
supersymmetric. However, we have observed above that this does not require  $\Pi _a ^{\rm bulk}$ to be $\mathcal{R}$-invariant, 
as $\Pi _{\rm O6}$ is. Thus, unlike  the $\mathbf{Z}_6$ case,   in this case $a \cap \Pi _{\rm O6}$ is generally non-zero.
 We noted in the Introduction that  
we must exclude the appearance of the representations ${\bf S}_a$ and ${\bf S}_b$. 
Consequently, we impose the constraints
\bea
a \cap a'&=& a \cap \Pi _{\rm O6} \label{aa'o6}\\
b \cap b'&=& b \cap \Pi _{\rm O6} \label{bb'o6}
\eea
We also showed that demanding that the $U(1)$ charges $Q_a$ and $Q_b$ sum to zero overall requires that 
$\#({\bf A}_a)=0=\#({\bf A}_b)$, 
at least if we also demand standard-model Yukawa couplings. However, for the moment we proceed more conservatively.
With the  constraint (\ref{aa'o6}) the  number of multiplets in the antisymmetric representation ${\bf A}_a$ is $a \cap \Pi _{\rm O6}$.  
For the present we require only that
\beq
 |a \cap \Pi _{\rm O6}|\leq 3  \label{aopio6}
 \eeq
  since otherwise there would again be non-minimal vector-like quark singlet matter. 
Similarly, using just (\ref{bb'o6}), we only require that
\beq
|b \cap \Pi _{\rm O6}|\leq 3
 \label{bopio6}
\eeq
to avoid unwanted vector-like lepton singlets.


\section{Results and conclusions}
We have shown \cite{Bailin:2006zf} that, unlike the $\mathbf{Z}_6$ orientifold,
 at least on some lattices, the   $\mathbf{Z}_6'$ orientifold 
{\em can} support  supersymmetric stacks $a$ and $b$ of D6-branes 
with intersection numbers satisfying $(a \circ b,a \circ b')=(2,1)$ or $(1,2)$. Stacks having this property are an 
indispensable ingredient in any intersecting brane model that has {\em just} the matter content of the (supersymmetric) 
standard model. 
 By construction, in all of our solutions  
there is no matter in symmetric representations of the gauge groups on either stack. However, some of the solutions {\em do} 
have matter,  2 quark  singlets $q^c_L$ or 2 lepton   singlets $\ell ^c_L$,  in  the antisymmetric representation  of gauge group
 on one of the stacks. This is not possible on the 
$\mathbf{Z}_6$ orientifold because all supersymmetric D6-branes wrap the same bulk 3-cycle as the O6-planes.
In contrast, on the  $\mathbf{Z}_6'$ orientifold there exist supersymmetric 3-cycles that do not wrap the O6-planes. Thus, 
there is more latitude in this case, and the solutions with antisymmetric matter exploit this feature. 
 Unfortunately, however, none of the solutions of this nature that we have found can be enlarged to give just the 
 standard-model spectrum, since the overall cancellation of the relevant $U(1)$ charge cannot be achieved with  this matter content.  
 Nevertheless, some of 
our solutions have no antisymmetric (or symmetric) matter on either stack. 
 We shall attempt in  a future work to construct  a  realistic (supersymmetric) standard model using one of these solutions.

The  presence of singlet matter on the branes in some, but not all,  of our solutions is an important feature of our results.
  It is clear that different orbifold point groups produce different physics, 
as indeed, for the reasons just given, our results also illustrate. The point group must act as an automorphism of the lattice used, but
 it is less clear that realising a given point group symmetry on 
different lattices produces different physics. Our results show that different lattices can produce different physics.
The 
observation that the lattice does affect the physics suggests that  other lattices are worth investigating in both the $\mathbf{Z}_6$ and 
$\mathbf{Z}_6'$ orientifolds. In particular, since $Z_6$ can be realised on a $G_2$ lattice, as well as  on an $SU(3)$ lattice, one or more of all 
 three $SU(3)$ lattices in the $\mathbf{Z}_6$ case, and of the two  on $T^2_{1,2}$ in the $\mathbf{Z}_6' $ case, could be replaced by a $G_2$ lattice. 
 We shall explore this avenue too in future work.
 
 The construction of a realistic model will, of course, entail adding further stacks of D6-branes $c,d,..$,  with just a single brane in 
 each stack, arranging that the matter content is just that of the supersymmetric standard model, the whole set 
 satisfying  
 the  condition 
for RR tadpole cancellation.
 In a supersymmetric orientifold,  RR tadpole cancellation ensures that NSNS tadpoles are also cancelled, 
  but some moduli, (some of)  of the complex structure moduli, the K\"ahler moduli 
 and the dilaton, remain unstabilised. Recent developments  have shown how such moduli may be stabilised using 
 RR, NSNS and metric fluxes \cite{Derendinger:2004jn,Kachru:2004jr,Grimm:2004ua,Villadoro:2005cu,DeWolfe:2005uu}, 
 and indeed C\'amara, Font \& Ib\'a\~nez \cite{Camara:2005dc, Aldazabal:2006up} have shown how  models 
 similar to the ones we have been discussing can be 
 uplifted into ones with stabilised K\"ahler moduli using a ``rigid corset''.
 In general,   such fluxes contribute to 
 tadpole cancellation conditions and might make them easier to satisfy. In which case, it may be that one or other of our 
 solutions with antisymmetric matter could be used to obtain just the standard-model spectrum. In contrast, the rigid corset 
 can be added to any RR tadpole-free assembly of D6-branes in order to stabilise all moduli. Thus our results represent 
 an important first step to obtaining a supersymmetric standard model from intersecting branes with all moduli stabilised.

\end{document}